\documentclass[twocolumn,prl,showpacs,superscriptaddress,preprintnumbers,amsmath,amssymb]{revtex4}

\usepackage[dvips]{graphicx}
\usepackage{dcolumn}
\usepackage{bm}
\begin{document}
%\preprint{APS/123-QED}
\title{Determination of the orbital moment and crystal field splitting in LaTiO$_{3}$}

\author{M. W. Haverkort}
 \affiliation{II. Physikalisches Institut, Universit{\"a}t zu K{\"o}ln,
  Z{\"u}lpicher Str. 77, D-50937 K{\"o}ln, Germany}
\author{Z. Hu}
 \affiliation{II. Physikalisches Institut, Universit{\"a}t zu K{\"o}ln,
  Z{\"u}lpicher Str. 77, D-50937 K{\"o}ln, Germany}
\author{A. Tanaka}
 \affiliation{Department of Quantum Matter, ADSM, Hiroshima University, Higashi-Hiroshima 739-8530, Japan}
\author{G. Ghiringhelli}
 \affiliation{INFM, Dipartimento di Fisica, Politecnico di Milano, p. Leonardo da Vinci 32, Milano 20133, Italy}
\author{H. Roth}
 \affiliation{II. Physikalisches Institut, Universit{\"a}t zu K{\"o}ln,
  Z{\"u}lpicher Str. 77, D-50937 K{\"o}ln, Germany}
\author{M. Cwik}
 \affiliation{II. Physikalisches Institut, Universit{\"a}t zu K{\"o}ln,
  Z{\"u}lpicher Str. 77, D-50937 K{\"o}ln, Germany}
\author{T. Lorenz}
 \affiliation{II. Physikalisches Institut, Universit{\"a}t zu K{\"o}ln,
  Z{\"u}lpicher Str. 77, D-50937 K{\"o}ln, Germany}
\author{C. Sch{\"u}{\ss}ler-Langeheine}
 \affiliation{II. Physikalisches Institut, Universit{\"a}t zu K{\"o}ln,
  Z{\"u}lpicher Str. 77, D-50937 K{\"o}ln, Germany}
\author{S. V. Streltsov}
 \affiliation{Institute of Metal Physics, S. Kovalevskoy 18, 620219 Ekaterinburg GSP-170, Russia}
\author{A. S. Mylnikova}
 \affiliation{Institute of Metal Physics, S. Kovalevskoy 18, 620219 Ekaterinburg GSP-170, Russia}
\author{V. I. Anisimov}
 \affiliation{Institute of Metal Physics, S. Kovalevskoy 18, 620219 Ekaterinburg GSP-170, Russia}
\author{C. de Nadai}
 \affiliation{European Synchrotron Radiation Facility, Bo\^{i}te Postale 220, Grenoble 38043, France}
\author{N. B. Brookes}
 \affiliation{European Synchrotron Radiation Facility, Bo\^{i}te Postale 220, Grenoble 38043, France}
\author{H. H. Hsieh}
 \affiliation{Chung Cheng Institute of Technology, National Defense University, Taoyuan 335, Taiwan}
\author{H.-J. Lin}
 \affiliation{National Synchrotron Radiation Research Center, 101 Hsin-Ann Road, Hsinchu 30077, Taiwan}
\author{C. T. Chen}
 \affiliation{National Synchrotron Radiation Research Center, 101 Hsin-Ann Road, Hsinchu 30077, Taiwan}
\author{T. Mizokawa}
 \affiliation{Department of Complexity Science and Engineering, University of Tokyo, Tokyo 113-0033, Japan}
\author{Y. Taguchi}
 \altaffiliation{\vspace{-5mm}Present address: Institute for Materials Research, Tohoku University, Sendai 980-8577, Japan}
 \affiliation{Department of Applied Physics, University of Tokyo, Tokyo, 113-8656, Japan}
\author{Y. Tokura}
 \affiliation{Department of Applied Physics, University of Tokyo, Tokyo, 113-8656, Japan}
 \affiliation{Spin Superstructure Project, ERATO, Japan Science and Technology Corporation, AIST Tsukuba Central 4, Tsukuba 305-8562, Japan}
\author{D. I. Khomskii}
 \affiliation{II. Physikalisches Institut, Universit{\"a}t zu K{\"o}ln,
  Z{\"u}lpicher Str. 77, D-50937 K{\"o}ln, Germany}
\author{L. H. Tjeng}
 \affiliation{II. Physikalisches Institut, Universit{\"a}t zu K{\"o}ln,
  Z{\"u}lpicher Str. 77, D-50937 K{\"o}ln, Germany}

\date{\today}

\begin{abstract}
Utilizing a sum-rule in a spin-resolved photoelectron
spectroscopic experiment with circularly polarized light, we show
that the orbital moment in LaTiO$_3$ is strongly reduced from its
ionic value, both below and above the N\'{e}el temperature. Using
Ti $L_{2,3}$ x-ray absorption spectroscopy as a local probe, we
found that the crystal field splitting in the $t_{2g}$ subshell
is about 0.12-0.30 eV. This large splitting does not facilitate
the formation of an orbital liquid.
\end{abstract}


\maketitle

LaTiO$_3$ is an antiferromagnetic insulator with a pseudocubic perovskite
crystal structure \cite{MacLean79,Eitel86,Cwik03}. The N\'{e}el temperature
varies between $T_N$ = 130 and 146 K, depending on the exact oxygen
stoichiometry \cite{Cwik03,Goral83,Meijer99}. A reduced total moment of about
0.45-0.57 $\mu_B$ in the ordered state has been observed
\cite{Cwik03,Goral83,Meijer99}, which could imply the presence of an orbital
angular momentum that is antiparallel to the spin momentum in the Ti$^{3+}$
$3d^1$ ion \cite{Meijer99,Mizokawa96}. In a recent Letter, however, Keimer {\it
et al.} \cite{Keimer00} have reported that the spin wave spectrum is nearly
isotropic with a very small gap, and concluded that therefore the orbital
moment must be quenched. To explain the reduced moment, they proposed the
presence of strong orbital fluctuations in the system. This seems to be
supported by the theoretical study of Khaliullin and Maekawa
\cite{Khaliullin00}, who suggested that LaTiO$_3$ is in an orbital liquid
state. If true, this would in fact constitute a completely novel state of
matter. By contrast, Cwik \textit{et al.} \cite{Cwik03}, Mochizuki and Imada
\cite{Mochizuki03}, as well as Pavarini \textit{et al.} \cite{Pavarini04}
estimated that small orthorhombic distortions present in LaTiO$_{3}$ would
produce a crystal field (CF) splitting strong enough to lift the Ti $3d$
$t_{2g}$ orbital degeneracy. However, one of the latest theoretical papers
finds a much smaller CF splitting, leaving open the possibility for an orbital
liquid state \cite{Solovyev04}.

The objective of our work is to establish via an independent
experimental method whether or not the orbital moment in
LaTiO$_{3}$ is quenched, and if so, whether the explanation for
the properties of LaTiO$_{3}$ should be searched in the proposed
orbital liquid picture or rather in terms of a large local CF
splitting. In other words, the magnitude of the CF splitting has
to be determined. We have carried out spin-resolved photoemission
(PES) experiments using circularly polarized light, and by
applying a sum-rule we have determined unambiguously that the
orbital moment is indeed strongly reduced from its ionic value in
a wide temperature range. We have also performed temperature
dependent Ti $L_{2,3}$ x-ray absorption (XAS) measurements, and
found from this local probe that the Ti $3d$ $t_{2g}$ orbitals
are split by about 0.12-0.30 eV. Our results are consistent with
the conclusion of Keimer \textit{et al.} in that the orbital
moment is very small. However, the sizable CF splitting does not
provide conditions favorable for the realization of an orbital
liquid.

Twinned single crystals of LaTiO$_{3}$ with $T_N$ = 146 K have
been grown by the traveling floating-zone method. The PES
experiments were performed at the ID08 beamline of the ESRF in
Grenoble. The photon energy was set to 700 eV, sufficiently high
to ensure bulk sensitivity \cite{Weschke91,Sekiyama00}. The
degree of circular polarization was close to $100\%$ and the spin
detector had an efficiency (Sherman function) of 17\%. The
combined energy resolution for the measurements was 0.6 eV and
the angle $\theta$ between the Poynting vector of the light and
the analyzer was 60$^\circ$. The XAS measurements were carried
out at the Dragon beamline of the NSRRC in Taiwan, with a photon
energy resolution set at 0.15 eV for the Ti $L_{2,3}$ edges
($h\nu \approx 450-470$ eV). The spectra were recorded using the
total electron yield method. Clean sample areas were obtained by
cleaving the crystals inside the measuring chambers with a
pressure of low 10$^{-10}$ mbar.

  \begin{figure}
    \includegraphics[width=0.45\textwidth]{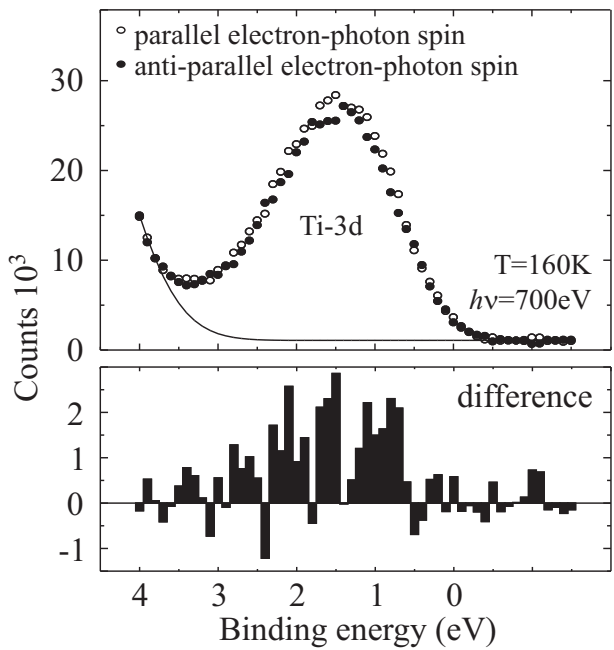}
    \caption{Spin-resolved photoemission spectra of twinned LaTiO$_{3}$
    single crystal taken with circularly polarized light.}
    \label{fig1}
  \end{figure}

Fig. 1 shows the spin-resolved photoemission spectra of the LaTiO$_3$ $3d$
states in the valence band, taken with circularly polarized light. The spectra
are corrected for the spin detector efficiency. One can observe a small but
reproducible difference between the spectra taken with the photon spin (given
by the helicity of the light) parallel or antiparallel to the electron spin.
The relevant quantity to be evaluated here is the integrated intensity of the
difference spectrum ($\int_{dif}$) relative to that of the integrated intensity
of the sum spectrum ($\int_{sum}$). This can be directly related to the
expectation value of the spin-orbit operator (\textbf{l.s}) applied to the
\textit{initial} state, thanks to the sum rule developed by van der Laan and
Thole \cite{Laan93}. For a $3d$ system in which the final states are mainly of
$f$ character due to the high photon energies used, and for a randomly oriented
sample, we obtain \cite{Ghiringhelli02}:
\begin{equation}
\frac{\int_{dif}}{\int_{sum}} = -U(\theta)
 \frac{\langle \sum_{i} \textbf{l}_{i} \cdot \textbf{s}_{i} \rangle }
{3 \langle n \rangle} \label{Eq1}
\end{equation}
where $U(\theta)$=$(3$-$4cos^{2}(\theta))$/$(2$-$cos^{2}(\theta))$ is a
geometrical factor to account for the angle between the Poynting vector of the
light and the outgoing photoelectron, the index \textit{i} runs over the
electrons in the $3d$ shell and $\langle n \rangle$ is the number of $3d$
electrons contributing to the spectra.

\begin{figure}
    \includegraphics[width=0.45\textwidth]{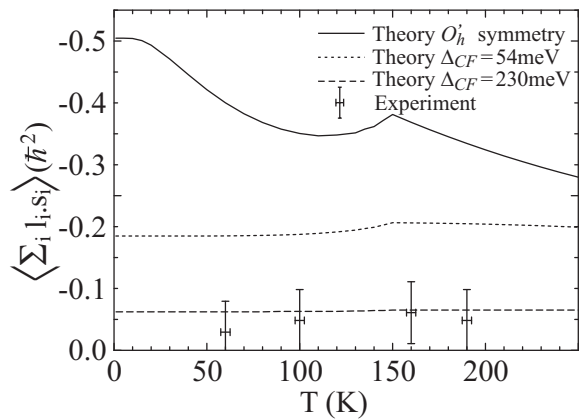}
    \caption{$\langle \sum_{i} \textbf{l}_{i} \cdot \textbf{s}_{i}
    \rangle$ values extracted from the spin-resolved circularly polarized
    photoemission data, together with theoretical predictions for various
    crystal field parameters.}
    \label{fig2}
  \end{figure}

With $\int_{dif}$/$\int_{sum}$$\approx$0.03, $\theta$=60$^\circ$, and $\langle
n \rangle$$\approx$0.8 from our cluster calculations \cite{Tanaka94}, we arrive
at $\langle \sum_{i} \textbf{l}_{i} \cdot \textbf{s}_{i} \rangle$$\approx$-0.06
(in units of $\hbar^2$), see Fig. 2. This is, in absolute value, an order of
magnitude smaller than the maximum possible value of -0.50 for a $3d^{1}$
$t_{2g}$ ion with $s_z$=1/2 and $l_z$=-1 (in units of $\hbar$). In fact, the
-0.06 value is so small, that we can directly conclude that for this $3d^{1}$
ion the orbital momentum is practically quenched. Fig. 2 shows that this is the
case for a wide range of temperatures, both below and above $T_{N}$.

Having established that LaTiO$_3$ has a strongly reduced orbital
moment, we now focus on the issue whether this is caused by
strong orbital fluctuations \cite{Keimer00,Khaliullin00} or
rather by strong local CF effects as theoretically proposed
\cite{Cwik03,Mochizuki03,Pavarini04}. To this end, we carry out
temperature dependent XAS measurements at the Ti $L_{2,3}$ ($2p
\rightarrow 3d$) edges. Here we make use of the fact that the
$2p$ core hole produced has a strong attractive Coulomb
interaction with the $3d$ electrons. This interaction is about 6
eV, and is more than one order of magnitude larger than the band
width of the $3d$ $t_{2g}$ states. The absorption process is
therefore strongly excitonic, making the technique an ideal and
extremely sensitive local probe \cite{Tanaka94,deGroot94,Thole97}.

 \begin{figure*}
    \includegraphics[width=0.9\textwidth]{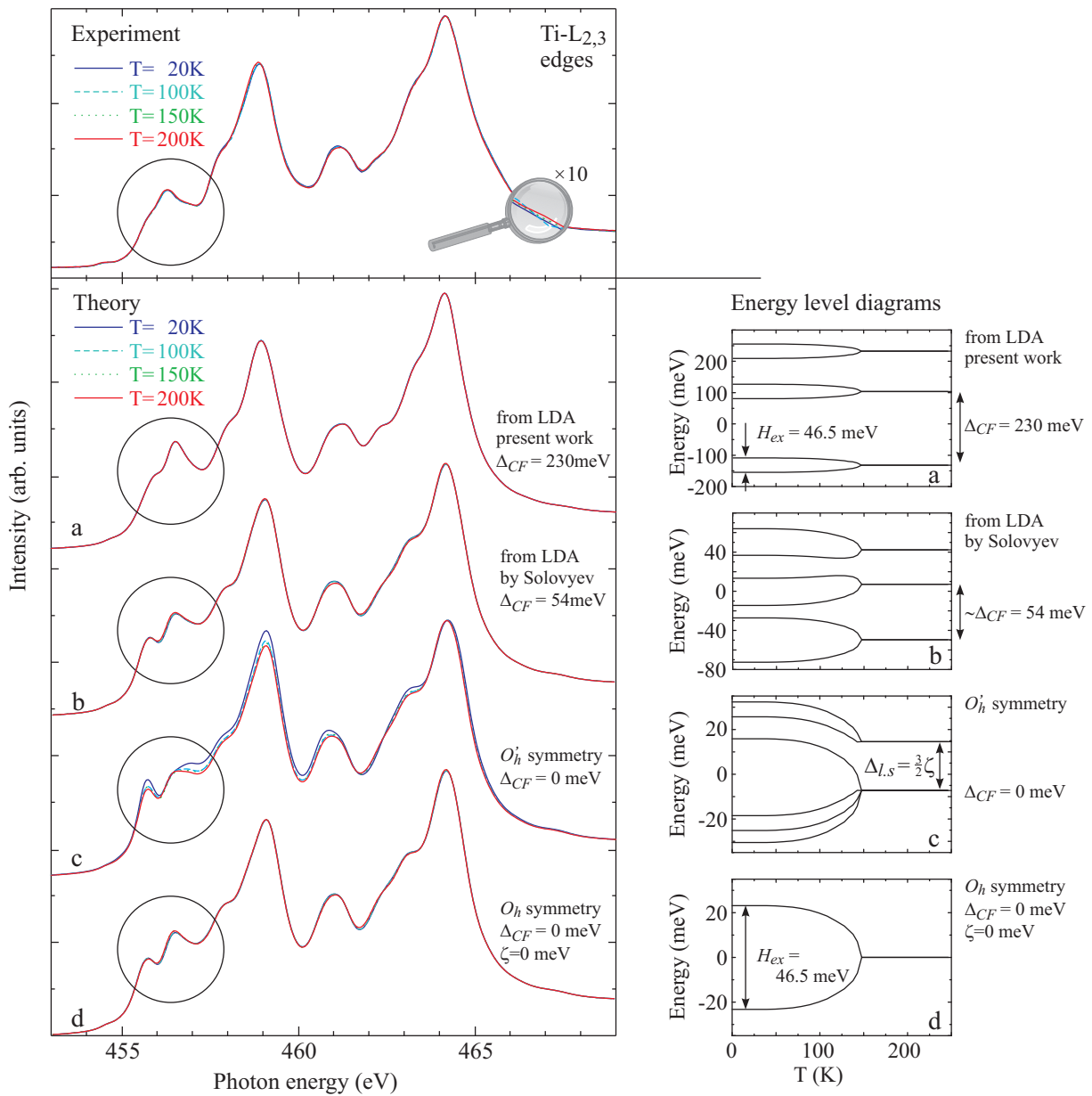}
    \caption{(Color online) Top panel: experimental Ti $L_{2,3}$ XAS spectra taken from
    a twinned LaTiO$_{3}$ single crystal at 20, 100, 150 and 200 K.
    Left panel: simulated isotropic spectra calculated for a TiO$_{6}$ cluster
    at 20, 100, 150 and 200 K for several CF parameters. Right panel:
    corresponding energy level diagrams for the cluster
    in an exchange field of $H_{ex}$ = 46.5 meV (from Keimer \textit{et al.}
    \cite{Keimer00}) at $T$ = 0 K and vanishing at $T_N$ = 146 K.
    Four scenarios are presented: (a) non-cubic symmetry with
    $\Delta_{CF}$ = 230 meV from our LDA calculation \cite{Streltsov04},
    (b) non-cubic symmetry with $\Delta_{CF}$ = 54 meV from
    Solovyev \cite{Solovyev04}, (c) $O_{h}^{'}$ and (d) $O_{h}$ symmetry.
    The spin-orbit constant $\zeta$ is 15.2 meV for (a), (b) and (c), and 0
    for (d). Note the very different energy scales.}
    \label{fig3}
  \end{figure*}

The top panel of Fig. 3 shows the experimental Ti $L_{2,3}$ XAS spectra for
several temperatures below and above $T_N$. One can clearly observe that the
spectra are temperature independent. In the subsequent sections we will discuss
two aspects of the spectra that are relevant for the determination of the
energetics and symmetry of the ground state and the lowest excited states of
LaTiO$_{3}$. The first is the detailed line shape of the spectra, and the
second is their temperature insensitivity.

To start with the first aspect, we have performed simulations in
order to obtain the best match with the experimental spectra, and
by doing so, to determine the magnitude of the CF splitting in
the $t_{2g}$ levels. For this we have used the well-proven
configuration interaction cluster model that includes the full
atomic multiplet theory and the hybridization with the O $2p$
ligands \cite{Tanaka94,deGroot94,Thole97}. Curves (a) in left
panel of Fig. 3 are the calculated isotropic spectra of a
TiO$_{6}$ cluster with a non-cubic crystal field splitting of
$\Delta_{CF}$ = 230 meV, as obtained, using a Wannier function
projection procedure, from our LDA calculation \cite{Streltsov04}
on the refined orthorhombic crystal structure \cite{Cwik03}. One
can see that the experimental data are well reproduced. We have
also carried out simulations with other $\Delta_{CF}$ values, and
found that $\Delta_{CF}$ should be in the range of about 120 to
300 meV in order to maintain the good agreement. If we chose, for
example, $\Delta_{CF}$ = 54 meV as proposed from the LDA
calculations by Solovyev \cite{Solovyev04}, we find that the
simulated line shapes are less satisfactory: curves (b) show
deviations from the experimental spectra, especially in the
encircled region. More important is that the situation without CF
splitting, i.e. in $O_{h}^{'}$ symmetry as shown by curves (c),
definitely does not agree with the experiment. Also the case as
depicted by curves (d), in which the spin-orbit interaction in
$O_{h}$ symmetry is artificially switched off as to obtain fully
degenerate $t_{2g}$ levels, which was the starting point of the
treatment of Khaliullin and Maekawa \cite{Khaliullin00}, does not
agree with the measurement. From the line shape analysis we can
thus firmly conclude that the crystal field splitting in
LaTiO$_{3}$ is quite appreciable.

\begin{figure}
    \includegraphics[width=0.45\textwidth]{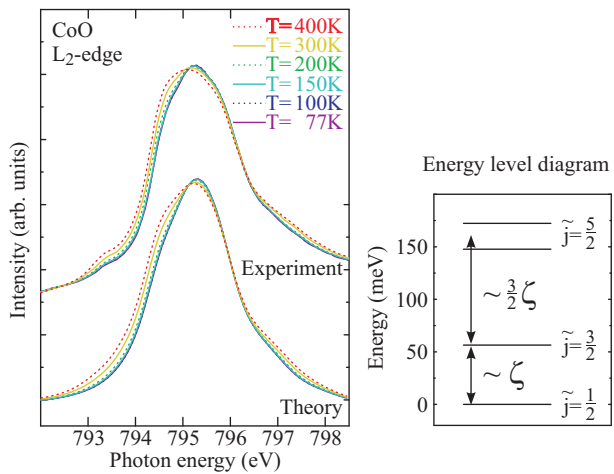}
    \caption{(Color online) Experimental temperature dependent Co $L_2$ XAS spectra of
    polycrystalline CoO, together with the simulated isotropic spectra
    and corresponding energy level diagram.}
    \label{fig4}
  \end{figure}

The second aspect of the Ti $L_{2,3}$ XAS spectra is their temperature
insensitivity. This may look like a trivial observation, but actually it is
not. For a $3d$ system with an open $t_{2g}$ shell, one usually expects to see
an appreciable temperature dependence in the isotropic spectrum: for instance,
in Fig. 4 we depict the Co $L_{2}$ XAS spectra of polycrystalline CoO, and
indeed, we do see a strong temperature dependence. The reason for this behavior
is that for a system with an unquenched orbital moment like CoO, the ground
state and the lowest excited states are split in energy by the spin-orbit
interaction and are separated in energy by an amount of the order of the
spin-orbit coupling \cite{deGroot93}. Since the final states that can be
reached from the ground state and from the lowest excited states are very
different, the spectrum will change with temperature depending on how much each
of the initial states is thermally populated. In Fig. 4 we have also simulated
the CoO spectra using a CoO$_{6}$ cluster model, and clearly the temperature
dependence is reproduced. In the left panel of Fig. 3, we have calculated the
LaTiO$_3$ spectra assuming a perfect $O_{h}^{'}$ local symmetry, and the
resulting curves (c) show indeed also a strong temperature dependence. However,
the fact that experimentally the LaTiO$_3$ spectra are temperature independent,
indicates directly that the spin-orbit interaction is inactive in LaTiO$_3$.
Indeed, simulations carried out for CF splittings much larger than the
spin-orbit interaction, e.g. curves (a) and (b) in Fig. 3, are not temperature
sensitive. The experimentally observed temperature insensitivity is therefore
fully consistent with the very small orbital moment found from the
spin-resolved photoemission measurements. We would like to note that the XAS
simulations were carried out including, for completeness, the presence of an
exchange field as depicted in the right panel of Fig. 3, although this had a
negligible influence on the isotropic spectra.

Returning to the spin-resolved photoemission data, we are able to
reproduce the very low $\langle \sum_{i} \textbf{l}_{i} \cdot
\textbf{s}_{i} \rangle$ of about -0.06 if we use $\Delta_{CF}$
values in the range of 120 and 300 meV. In Fig. 2 we show the
results calculated for the $\Delta_{CF}$ = 230 meV as found from
our LDA. The corresponding extracted orbital moment is $L_{z}$ =
-0.06. The $\Delta_{CF}$ = 54 meV value as proposed by Solovyev
\cite{Solovyev04}, however, clearly gives a $\langle \sum_{i}
\textbf{l}_{i} \cdot \textbf{s}_{i} \rangle$ value that deviates
substantially from the experimental one. The orbital moment in
this scenario is quite large: $L_{z}$ = -0.24. It is almost
superfluous to note that the calculation with $\Delta_{CF}$ = 0
meV, i.e. in perfect $O_{h}^{'}$ symmetry, gives results that are
in strong disagreement with the experiment.

To conclude, we have observed that the orbital moment in
LaTiO$_3$ is strongly reduced from its ionic value, supporting the
analysis from the neutron experiment by Keimer {\it et al.}
\cite{Keimer00}. Our experiments have also revealed the presence
of non-cubic crystal fields sufficiently strong to split the Ti
$t_{2g}$ levels by about 0.12-0.30 eV, confirming several of the
theoretical estimates
\cite{Cwik03,Mochizuki03,Pavarini04,Streltsov04}. Such a large
crystal field splitting provides a strong tendency for the Ti
$3d$ orbitals to be spatially locked, i.e. the quadrupole moment
measured at 1.5 K by NMR \cite{Kiyama03} should also persist at
the more relevant higher temperatures, making the formation of an
orbital liquid in LaTiO$_{3}$ rather unfavorable.

We acknowledge Lucie Hamdan for technical assistance. The research in K\"oln is
supported by the Deutsche Forschungsgemeinschaft through SFB 608 and the
research in Ekaterinburg by grants RFFI 04-02-16096 and yp.01.01.059.

\end{document}